\documentclass[11pt]{article}
\usepackage{amsmath,amssymb,amsfonts}
\usepackage[english]{babel}
\makeatletter
\@addtoreset{equation}{section}
\makeatother

\def\nn{\nonumber} \def\bd{\begin{document}} \def\ed{\end{document}}
\def\ds{\documentstyle}
\let\bm=\bibitem
\newcommand{\be}{\begin{equation}}
\newcommand{\ee}{\end{equation}}
\newcommand{\bea}{\setlength\arraycolsep{2pt} \begin{eqnarray}}
\newcommand{\eea}{\end{eqnarray}}
\newcommand{\hoch}[1]{$\, ^{#1}$}
\def\p{\partial}

\title{\large {\bf A note on the mass of Kerr-AdS black holes in the off-shell
generalized ADT formalism}}
\date{}

\author{Yi-De Jing and Jun-Jin Peng\footnote{Corresponding author: pengjjph@163.com}  \\ \\
\small \sl School of Physics and Electronic Science, Guizhou Normal University,\\
\small Guiyang, Guizhou 550001, People's Republic of China
}

\begin{document}
\maketitle
\vspace{20pt}

\begin{abstract}
In this note, the off-shell generalized Abbott-Deser-Tekin (ADT) formalism is applied to
explore the mass of Kerr-anti-de Sitter (Kerr-AdS) black holes in various dimensions
within asymptotically rotating frames. The cases in four and five dimensions are
explicitly investigated. It is demonstrated that the asymptotically rotating effect
may make the charge be non-integrable or unphysical when the asymptotic
non-rotating timelike Killing vector associated with the charge is allowed to vary
and the fluctuation of the metric is determined by the variation of all the mass
and rotation parameters. To avoid such a dilemma, we can let the
non-rotating timelike Killing vector be fixed or perform calculations in the
asymptotically static frame.
Our results further support that the ADT formalism is background-dependent.
\end{abstract}



\newpage
\voffset=-.90pt
\vspace{40pt}


\section{Introduction}

Till now, exact rotating black hole solutions with cosmological constant have been
found in various dimensions within the context of Einstein gravity.
In 1968, Carter first found a generalization of the four-dimensional (4D) rotating Kerr
black hole with a cosmological constant \cite{4DKeAdSsolu}. Since this black hole has
asymptotically de Sitter (dS) or anti-de Sitter (AdS) boundary conditions, it is
usually called as Kerr-dS or Kerr-AdS black hole in the literature. Many years later,
Hawking, Hunter and Taylor-Robinson found the five-dimensional (5D) generalization of
the 4D Kerr-(A)dS black hole, as well as the solutions with just one nonzero
angular momentum parameter in all dimensions \cite{5DsoluHHT}. In fact, the 5D Kerr-(A)dS black
hole can also be regarded as a generalization of the 5D Ricci-flat rotating Myers-Perry
black hole \cite{MyersP} including a cosmological constant. Subsequently, Gibbons, L\"{u}, Page
and Pope further constructed the general Kerr-(A)dS black holes with arbitrary
angular momenta in all higher dimensions \cite{GLuPPA,GLuPPB}, which exactly satisfy the
vacuum Einstein field equation with a cosmological constant. For the Kerr-AdS black holes,
inspired by string theory, especially by the anti-de Sitter/conformal field theory
(AdS/CFT) correspondence, as well as black hole thermodynamics, much work has been
done on their diverse aspects
\cite{BMajhiLamd,McIOn,TsaWY,SQWsepa,AshPaBr,CaiCaoP,NashCPB,ObukRub,CaiCaoC,PaPasken,HajShe,
Olen4Dker,JamsarDker,GibPPther,DerKather,GullTek,BarnichC,HollIshM,ArosCOTZ,AMDmassA,AMDmassB}.

Recently, in Ref. \cite{KimKY}, relieving the restriction that the background spacetime
satisfies the field equations, Kim, Kulkarni and Yi proposed a quasi-local formalism of
conserved charges within the framework of generic covariant pure gravity theories by
constructing an off-shell ADT current to generalize the conventional on-shell Noether
potential in the usual ADT formulation \cite{AbbottDA,AbbottDB,DeserTA,DeserTB} to the off-shell level, as well as following
the Barnich-Brandt-Compere (BBC) method \cite{BarnichB,Barnich,BarnichC} to incorporate a single
parameter path in the space of solutions into their definition. Since the current and
potential in the modified approach is off-shell, one may refer to it as the off-shell
generalized ADT formalism. In contrast with the usual one, the off-shell generalization makes it more
operable to derive the Noether potential from the corresponding current
and the procedure of computation become more convenient to manipulate. Owing to these,
it provides another fruitful way to evaluate the ADT charges for various theories
of gravity. For several developments and applications of the off-shell generalized
ADT formalism see the works
\cite{CiteHJPY,JJPeng,JJPengpform,HJPYLamb,SetAdam,JJPXcai,JJPIJMPA,LiLuWei,SWuLi}.

As usual, after obtaining the Kerr-AdS black hole solutions, it is of great necessity
to identify their mass and angular momenta. Because of the asymptotically AdS structure,
the usual Arnowitt-Deser-Misner (ADM) formalism, as well as the Komar integral, fails
to produce their mass. So it is desired to seek for other feasible approaches.
Fortunately, some works
\cite{NashCPB,ObukRub,CaiCaoC,PaPasken,HajShe,Olen4Dker,JamsarDker,GibPPther,
DerKather,GullTek,BarnichC,HollIshM,ArosCOTZ,AMDmassA,AMDmassB} have succeeded to yield the conserved
charges of the Kerr-AdS black holes through a series of methods, such as the
Ashtekar-Magnon-Das (AMD) formalism \cite{AMDmassA,AMDmassB}, the (off-shell) ADT formulation,
the BBC method and so on. For example, in the work \cite{GullTek}, the usual ADT
formulation has been applied to obtain the mass of Kerr-AdS black holes in various
dimensions. To realise this, the perturbation of the metric is set as the divergence
between the metric and a fixed reference background, which is the spacetime got
through letting the parameter associated with the mass in
the original metric be zero. In \cite{BarnichC}, the same background was also adopted to
calculate the conserved charges of the Kerr-AdS black holes via the BBC method.
There the fluctuation of the metric only depends on the parameter related to
the mass rather than all the solution parameters. Besides, the potential used to define
the conserved charges coincides with that in the usual ADT formulation. Due to these,
the BBC method adopted in \cite{BarnichC} is essentially in accordance with
the usual ADT one.

Notably, the usual ADT formulation and the BBC approach are background-dependent.
Actually, in the works \cite{GullTek,BarnichC}, the reference spacetimes adopted to
calculate the mass of Kerr-AdS black holes in all dimensions are non-rotating at
infinity. Otherwise, both the two methods may fail to yield physical results if the
timelike Killing vector associated with the mass is chosen as the usual one
$\xi^\mu=-\delta^\mu_t$.
For the off-shell generalized ADT formalism, we also wonder what will happen if it is
used to deal with the mass of the Kerr-AdS black holes in an asymptotic rotating
frame. On the other hand, in the light of the fact that the mass together with
the angular momenta enters into the first law of thermodynamics for black holes
as thermodynamical variables, the rotation parameters should be the members to
determine the fluctuation of the metric for the Kerr-AdS black holes. However, in Ref.
\cite{BarnichC}, only the mass parameter was regarded as a variable to fluctuate
the metric, while the rotation parameters were fixed. In view of
above-mentioned issues, unlike the works \cite{GullTek,BarnichC}, we shall utilize the off-shell
generalized ADT formalism to explore the mass of the Kerr-AdS black holes in a more
general manner. Namely, we do this under the conditions that the background spacetime
is asymptotically rotating and the fluctuation of the metric is determined by the
variation of all the mass and rotation parameters. The results demonstrate that the way
that the Kerr-AdS black holes behave at infinity, rotating or not, plays a key role in
determining whether the off-shell generalized ADT formalism can successfully produce
their physically meaningful mass.

The remainder of this work goes as follows. Sections \ref{sectwo}  and \ref{secthree}
are devoted to investigating the off-shell generalized ADT mass of the Kerr-AdS black holes
in four and five dimensions respectively. In Section \ref{secfour}, the mass of the Kerr-AdS
black holes in arbitrary dimensions is calculated by generalizing the results in
the 4D and 5D cases. The last section is our conclusions.

\section{Mass of the 4D Kerr-AdS black hole via the off-shell generalized
ADT formalism}\label{sectwo}

As is well-known, the theory of Einstein gravity in $D$ dimensions is described by
the Einstein-Hilbert Lagrangian
\be
\mathcal{L}_{EH}=\sqrt{-g}\big(R -2\Lambda\big)
\,  \label{EHLagran}
\ee
with $\Lambda =-(D-1)(D-2)\ell^2/2$, where $\ell^{-1}$ is the radiu of AdS spaces.
The field equation for the gravitational field is
\be
R_{\mu\nu}=-\frac{(D-1)(D-2)}{2}\ell^2g_{\mu\nu}
\, . \label{MotionEq}
\ee
The 4D Kerr-AdS black hole \cite{4DKeAdSsolu} is an exact rotating solution with
asymptotic AdS behavior of Eq. (\ref{MotionEq}) in the case $D=4$. In Boyer-Lindquist
coordinates, the metric for the 4D Kerr-AdS black hole takes the form
\bea
ds_{(4)}^2&=& -\frac{\Delta_{(4)}}{\Sigma_{(4)}}\Big[dt-a\sin^2\theta
\Big(\frac{d\phi}{\Xi} - \omega_\phi \frac{dt}{\Xi}\Big)\Big]^2
+\frac{\Sigma_{(4)}}{\Delta_{(4)}}dr^2+\frac{\Sigma_{(4)}}{F_{(4)}} d\theta^2
\nn \\
&&+\frac{F_{(4)}\sin^2\theta}{\Sigma_{(4)}}\Big[adt-(r^2+a^2)
\Big(\frac{d\phi}{\Xi} - \omega_\phi \frac{dt}{\Xi}\Big)\Big]^2
\, , \label{Metric4DKeAdS}
\eea
in which
\bea
\Delta_{(4)}&=&(r^2+a^2)(1+\ell^2r^2)-2mr \, , \quad
\Sigma_{(4)}=r^2+a^2\cos^2\theta \, ,
\nn \\
F_{(4)}&=&1-a^2\ell^2\cos^2\theta \, , \quad
\Xi = 1-\ell^2 a^2
\, , \label{Func4DKerAdS}
\eea
and the constant $\omega_\phi=\omega_\phi(m,a,\ell)$, only depending on the integral parameters
$m$ and $a$, as well as the constant $\ell$. Particularly, when $\omega_\phi=0$, the metric
(\ref{Metric4DKeAdS}) becomes
the usual form of the 4D Kerr-AdS black hole, which is asymptotic to $AdS_4$ in a rotating
frame with the angular velocity $\Omega^\infty_\phi =-a\ell^2$. To guarantee that the black
hole is static at infinity, observed relative to a frame that is non-rotating, one only
needs to set $\omega_\phi=a\ell^2$.

We now go on to compute the mass of the 4D Kerr-AdS black hole via the off-shell generalized ADT
method proposed in \cite{KimKY}. With the choice of a Killing vector $\xi^\mu$, the definition of the ADT
conserved charge related to the Einstein-Hilbert Lagrangian (\ref{EHLagran}) is read off as
\be
\delta Q_c=\frac{1}{16\pi(D-2)!} \int_{\partial\Sigma} \sqrt{-g}\mathcal{Q}_{ADT}^{\mu\nu}
\epsilon_{\mu\nu\mu_1\mu_2\cdot\cdot\cdot\mu_{(D-2)}}dx^{\mu_1}\wedge\cdot\cdot\cdot
\wedge dx^{\mu_{(D-2)}}
\, , \label{dQdefineAn}
\ee
where the quantity $\epsilon_{\mu\nu\mu_1\mu_2\cdot\cdot\cdot\mu_{(D-2)}}$ is the totally
antisymmetric Levi-Civita tensor, which is defined through the equation
$\epsilon_{\mu_1\mu_2\cdot\cdot\cdot\mu_{D}}=
D!\delta^0_{[\mu_1}\delta^1_{\mu_2}\cdot\cdot\cdot\delta^{D-1}_{\mu_D]}$, and
the off-shell ADT potential is defined by
\bea
\mathcal{Q}_{ADT}^{\mu\nu}&=&Q_{ADT}^{\mu\nu}+\nabla^{[\mu}\delta\xi^{\nu]}
\, , \nn \\
Q_{ADT}^{\mu\nu}&=&\xi_\sigma \nabla^{[\mu}h^{\nu]\sigma}
-h^{\sigma[\mu}\nabla_\sigma\xi^{\nu]}
+\frac{1}{2}h\nabla^{[\mu}\xi^{\nu]}
-\xi^{[\mu}\nabla_\sigma h^{\nu]\sigma}
+\xi^{[\mu}\nabla^{\nu]}h
\, , \label{ADTpotential}
\eea
in which $h_{\mu\nu}=\delta g_{\mu\nu}$, $h=g^{\mu\nu}h_{\mu\nu}$, and $Q_{ADT}^{\mu\nu}$
is the conventional (off-shell) ADT potential. The term with $\delta\xi^\mu$ comes from the
variation of the off-shell Noether potential in accordance with the generalized
off-shell ADT potential in \cite{JJPengpform}, whose contribution to the off-shell ADT current
is $\nabla_\nu\nabla^{[\mu}\delta\xi^{\nu]}=R^\mu_\nu \delta\xi^\nu
+\frac{1}{2}\mathcal{L}_\xi \Theta^\mu$, where the surface term
$\Theta^\mu=2g^{\mu[\rho}\nabla_\nu h^{\nu]}_{\rho}$. Let us pay attention to the behaviour
of the charge in the case where the Killing vector $\xi^\mu$ is assumed to be fixed,
that is $\delta\xi^\mu=0$. In such a case, it has been shown that the charge $Q_c$
associated with the linear combination of two Killing vectors preserves the linearity
property in \textbf{Appendix A}. However, for the other cases where $\delta\xi^\mu\neq 0$,
the linear property of the charge may break down because of the appearance of terms
proportional to the Komar integral.

For the 4D Kerr-AdS black hole
described by the metric (\ref{Metric4DKeAdS}), the fluctuation of the spacetime
$h_{\mu\nu}$ is determined by the infinitesimal change of both the parameters
$(m,a)$ rather than the single mass parameter $m$ like in \cite{BarnichC},
that is
\be
m\rightarrow m+dm \, , \qquad
a\rightarrow a+da
\, . \label{machfor4DKeAdS}
\ee
Under such conditions, we calculate the charge associated with the 4D timelike
Killing $\xi_{(t)}^\mu=(-1,0,0,0)$, which is rotating at radial infinity.
The $(t,r)$ component of the off-shell ADT potential is computed as
\bea
\sqrt{-g}\mathcal{Q}_{ADT}^{tr}[\xi^\mu_{(t)}]
&=&\Upsilon_{(4)}r+\mathcal{O}\Big(\frac{1}{r}\Big)
+\sin\theta(2\Xi+3a\omega_\phi\sin^2\theta)\frac{dm}{\Xi^2}  \nn \\
&&+3m\sin\theta[2a\ell^2\Xi+(4-3\Xi)\omega_\phi\sin^2\theta]\frac{da}{\Xi^3}
\, , \nn \\
\Upsilon_{(4)}&=&-a\ell^2\sin\theta(1-3\cos^2\theta)\frac{da}{\Xi}
\, . \label{Qtrfor4DBH}
\eea
Substituting the above equation into the formula (\ref{dQdefineAn}) for the
conserved charge, together with the condition that the $\Upsilon_{(4)}$ term
makes no contribution to the mass since $\int^\pi_0\Upsilon_{(4)} d\theta =0$,
we have
\be
dQ_c[\xi^\mu_{(t)}]=\frac{1}{\Xi^3}\Big\{\Xi dm+4ma\ell^2da
+\big(\omega_\phi-a\ell^2\big)\big[a\Xi dm+m(4-3\Xi)da\big]\Big\}
\, . \label{dQ4DKerAdS1}
\ee
A straightforward computation then shows that
\be
dQ_c[\xi^\mu_{(t)}]=dM_{(4)}+\big(\omega_\phi+\Omega^\infty_\phi\big)dJ_{(4)}
\, , \label{dM4DKerAdSt}
\ee
in which, the quantities $M_{(4)}$ and $J_{(4)}$ are read off as
\be
M_{(4)}=\frac{m}{\Xi^2} \, , \qquad
J_{(4)}=\frac{ma}{\Xi^2}
\, , \label{MassAnfor4DBH}
\ee
respectively. They are the usual mass and angular momentum of the 4D Kerr-AdS black
hole presented in the literature
\cite{PaPasken,HajShe,Olen4Dker,JamsarDker,GibPPther,DerKather,GullTek,BarnichC}.
On the other hand, the charge $Q_c[\xi^\mu_{(\phi)}]$ associated with the spacelike
Killing vector $\xi_{(\phi)}^\mu=(0,0,0,1)$ is computed as
\be
dQ_c[\xi^\mu_{(\phi)}]=dJ_{(4)}
\, , \label{dQ4DKeranguM}
\ee
from which one can obtain the angular momentum $J_{(4)}$. In fact, it essentially
results from the Komar integral in Eq. (\ref{KomarInt}).

Since the 4D Kerr black hole is rotating at infinity, the Killing vector corresponding
to its mass should be chosen as the asymptotic non-rotating timelike Killing vector
$\hat{\xi}^\mu_{(t)}=\xi^\mu_{(t)}-\big(\omega_\phi+\Omega^\infty_\phi\big)\xi^\mu_{(\phi)}$,
which is just the linear combination of the Killing vector $\xi^\mu_{(t)}$ and
$\xi^\mu_{(\phi)}$. With the help of Eq. (\ref{DQc1c2}) in \textbf{Appendix A}, we have
\be
d\widetilde{M}_{(4)}=dM_{(4)}
-J_{(4)}\Big(\frac{\partial\big(\omega_\phi+\Omega^\infty_\phi\big)}{\partial m}dm
+\frac{\partial\big(\omega_\phi+\Omega^\infty_\phi\big)}{\partial a}da\Big)
\, , \label{dM4DKerAdS}
\ee
where $Q_c[\hat{\xi}^\mu_{(t)}]=\widetilde{M}_{(4)}$. Obviously, for a general
$\omega_\phi$, the second term in the right side of Eq. (\ref{dM4DKerAdS}),
arising from the variation of the Killing vector $\hat{\xi}^\mu_{(t)}$, may make
$\widetilde{M}_{(4)}$ be non-integrable in the case where both the parameters
$m$ and $a$ are variables when the angular velocity $(\omega_\phi+\Omega^\infty_\phi)\neq 0$.
In contrast to this, in the case where the metric perturbation
$h_{\mu\nu}$ merely depends on the change of the mass parameter $m$, like in
\cite{BarnichC}, $\widetilde{M}_{(4)}$ is integrable. Nevertheless, it is not
``physically meaningful" unless $(\omega_\phi+\Omega^\infty_\phi)$ vanishes or
$\omega_\phi$ is independent on the mass parameter. Therefore, in order to
guarantee that the mass $\widetilde{M}_{(4)}=M_{(4)}$, a rather efficient way
is to let the angular velocity at infinity vanish, namely,
$\omega_\phi=-\Omega^\infty_\phi=a\ell^2$, or to fix the Killing vector
$\hat{\xi}^\mu_{(t)}$, although such conditions are not required when the
off-shell generalized ADT method is applied to compute the angular momentum of
the 4D Kerr-AdS black hole.

\section{Mass of the 5D Kerr-AdS black hole}\label{secthree}

The 5D Kerr-AdS black hole, which behaves like an asymptotic $AdS_5$ space and
possesses two independent rotations, is an exact solution of the field equation
(\ref{MotionEq}) in five dimensions. This black hole was first constructed in
\cite{5DsoluHHT}, and it can be seen as a special case of the general solutions
in arbitrary higher dimensions, found in \cite{GLuPPA,GLuPPB}. For the sake of convenience
on our analysis, the metric for the 5D Kerr-AdS black hole takes the form
\bea
ds^2_{(5)}&=& -\frac{\Delta_{(5)}}{\Sigma_{(5)}}\Big[dt-a(1-x^2) d\hat{\phi}
-bx^2 d\hat{\psi}\Big]^2
+\frac{\Sigma_{(5)}}{\Delta_{(5)}}dr^2+\frac{\Sigma_{(5)}}{F_{(5)}}\frac{ dx^2}{1-x^2}
\nn \\
&&+\frac{1+\ell^2 r^2}{r^2\Sigma_{(5)}}\Big[abdt-b(1-x^2)(r^2+a^2) d\hat{\phi}
-ax^2(r^2+b^2) d\hat{\psi}\Big]^2
\nn \\
&&+\frac{F_{(5)}}{\Sigma_{(5)}}\big(1-x^2\big)\Big[adt-(r^2+a^2) d\hat{\phi} \Big]^2
+\frac{F_{(5)}}{\Sigma_{(5)}}x^2\Big[bdt-(r^2+b^2) d\hat{\psi} \Big]^2
\, , \label{Metricof5DBH}
\eea
where
\bea
\Delta_{(5)} &=&\frac{(r^2+a^2)(r^2+b^2)(1+\ell^2r^2)}{r^2}-2m
\, , \quad
\Sigma_{(5)}=r^2+a^2x^2+b^2(1-x^2) \, ,
\nn \\
F_{(5)}&=&1-a^2\ell^2x^2-b^2\ell^2(1-x^2) \, , \quad
\Xi_a=1-a^2\ell^2 \, , \quad \Xi_b=1-b^2\ell^2 \, ,
\nn \\
d\hat{\phi}&=&\frac{d\phi}{\Xi_a}-\omega_\phi\frac{dt}{\Xi_a} \, , \quad
d\hat{\psi}=\frac{d\psi}{\Xi_b}-\omega_\psi\frac{dt}{\Xi_b}
\, . \label{funcfor5DBH}
\eea
In Eq. (\ref{Metricof5DBH}), both the coordinates $\phi$ and $\psi$ range from
$[0, 2\pi]$, while $x$ takes the value $[0, 1]$. Significantly, when $\omega_\phi$
and $\omega_\psi$, which depend on  the four parameters $(m,a,b,\ell)$, disappear
and the coordinate $x$ is substituted by a new one $\theta$
through the relation $x=\cos\theta$, the metric (\ref{Metricof5DBH}) returns to
the usual form of the 5D Kerr-AdS black hole in the literature, whose angular
velocities along the $\phi$ and $\psi$ directions are $\Omega^\infty_\phi =-a\ell^2$
and $\Omega^\infty_\psi =-b\ell^2$ respectively at infinity. Unlike this, in the
more general cases $\omega_\phi, \omega_\psi\neq 0$, the angular velocities of
the metric (\ref{Metricof5DBH}) at infinity become $\omega_\phi+\Omega^\infty_\phi$
and $\omega_\psi+\Omega^\infty_\psi$.

In order to calculate the off-shell generalized ADT mass of the 5D Kerr-AdS black
hole (\ref{Metricof5DBH}), the fluctuation of the metric is determined by the
infinitesimal variation of the parameters $(m,a,b)$ as
\be
m\rightarrow m+dm \, , \quad
a\rightarrow a+da \, , \quad
b\rightarrow b+db \, , \quad
\ee
respectively. As before, we first consider the conserved charge associated with the Killing
vector $\xi^\mu_{(5t)} = (-1,0,0,0,0)$. In the light of the off-shell ADT
potential $\mathcal{Q}_{ADT}^{\mu\nu}$ in Eq. (\ref{ADTpotential}), a complex
calculation gives the following expression for the $(t,r)$ component of the
potential:
\bea
\sqrt{-g}\mathcal{Q}_{ADT}^{tr}[\xi^\mu_{(5t)}]&=&
\frac{1}{\Xi_a\Xi_b}\big[3xdm-2b(2+\Xi_b-\Xi_a)\big(2x^3-x\big)db\big]
\nn \\
&&+2X_{(5)}+4\omega_\phi \big(x-x^3\big)Y_{(5)}
+2X_{(5)}(a\leftrightarrow b)
\nn \\
&&+4\omega_\psi x^3Y_{(5)}(a\leftrightarrow b)
+\Upsilon_{(5)}r^2+\mathcal{O}\Big(\frac{1}{r^2}\Big)
\, , \label{Qtrfor5DKeBH}
\eea
in which
\bea
X_{(5)}&=&\frac{axda}{\Xi^2_a\Xi_b}\big[3\Xi_a(\Xi_a-\Xi_b)x^4+2\Xi_a(1+\Xi_b-\Xi_a)x^2
-\Xi_a+4m\ell^2\big] \, , \nn \\
Y_{(5)}&=&\frac{1}{\Xi^3_a\Xi^2_b}\big[a\Xi_a\Xi_b dm+m\Xi_b(4-3\Xi_a)da+2abm\Xi_a\ell^2db\big]
\, , \nn \\
\Upsilon_{(5)}&=&\frac{2\ell^2}{\Xi_a\Xi_b}\big(2x^3-x\big)\big(ada-bdb\big)
\, . \label{UpsiXY}
\eea
In the above equation, the quantity $\Upsilon_{(5)}$ has the property
$\int_0^1 \Upsilon_{(5)} dx=0$, which implies that the contribution to the mass
from the first term in Eq. (\ref{Qtrfor5DKeBH}) can be neglected. Another avenue to
cancel the contribution from the $r^2$ term is to let the perturbation of the metric
be independent of the rotation parameters $a$ and $b$, like in \cite{BarnichC}.
This holds for the cancellation of the contribution from the $r$ term in
Eq. (\ref{Qtrfor4DBH}) as well.
Further making use of the definition (\ref{dQdefineAn}) for the off-shell ADT
conserved charges, one gets
\be
dQ_c[\xi^\mu_{(5t)}]
=dM_{(5)}+\big(\omega_\phi+\Omega^\infty_\phi\big)dJ_{(5\phi)}
+\big(\omega_\psi+\Omega^\infty_\psi\big)dJ_{(5\psi)}
\, , \label{dM5DKerAdSt}
\ee
where $M_{(5)}$, $J_{(5\phi)}$ and
$J_{(5\psi)}$ are the mass and angular momenta along the
$\phi$ and $\psi$ directions got through other methods in the literature
\cite{JamsarDker,GibPPther,DerKather,GullTek,BarnichC}, which are read off as
\bea
M_{(5)}&=&\frac{\pi}{4}\frac{m}{\Xi^2_a\Xi^2_b}(2\Xi_a+2\Xi_b-\Xi_a\Xi_b)
\, , \nn \\
J_{(5\phi)}&=&\frac{\pi}{2}\frac{ma}{\Xi^2_a\Xi_b} \, , \quad
J_{(5\psi)}=\frac{\pi}{2}\frac{mb}{\Xi_a\Xi^2_b}
\, . \label{CCof5DKerrAdS}
\eea
By using the off-shell generalized ADT formula (\ref{dQdefineAn}), the angular
momenta $J_{(5\phi)}$ and $J_{(5\psi)}$, associated with the Killing
vectors $\xi^\mu_{(5\phi)}=\delta^\mu_\phi$ and $\xi^\mu_{(5\psi)}=\delta^\mu_\psi$
respectively, can also be obtained through
\be
dQ_c[\xi^\mu_{(5\phi)}]=dJ_{(5\phi)}\, , \quad
dQ_c[\xi^\mu_{(5\psi)}]=dJ_{(5\psi)}
\, . \label{5DannguMom}
\ee
It was explicitly proved in \cite{GibPPther} that the conserved charges $M_{(5)}$,
$J_{(5\phi)}$ and $J_{(5\psi)}$ strictly satisfy the first law of thermodynamics.
What is more, by making use of Eqs. (\ref{dM5DKerAdSt}), (\ref{5DannguMom}) and
(\ref{DQc1c2}), the off-shell ADT charge of the 5D Kerr-AdS black hole
$Q_c[\hat{\xi}^\mu_{(5t)}]$, where the asymptotic non-rotating timelike Killing vector
$\hat{\xi}^\mu_{(5t)}=\xi^\mu_{(t)}-\big(\omega_\phi+\Omega^\infty_\phi\big)\xi^\mu_{(5\phi)}
-\big(\omega_\psi+\Omega^\infty_\psi\big)\xi^\mu_{(5\psi)}$, is computed as
\bea
d\widetilde{M}_{(5)}&=&dM_{(5)}
-J_{(5\phi)}\Big(\frac{\partial\big(\omega_\phi+\Omega^\infty_\phi\big)}{\partial m}dm
+\frac{\partial\big(\omega_\phi+\Omega^\infty_\phi\big)}{\partial a}da
+\frac{\partial\big(\omega_\phi+\Omega^\infty_\phi\big)}{\partial b}db\Big)
\nn \\
&&-J_{(5\psi)}\Big(\frac{\partial\big(\omega_\psi+\Omega^\infty_\psi\big)}{\partial m}dm
+\frac{\partial\big(\omega_\psi+\Omega^\infty_\psi\big)}{\partial a}da
+\frac{\partial\big(\omega_\psi+\Omega^\infty_\psi\big)}{\partial b}db\Big)
\, , \label{dM5DKerAdS}
\eea
in which, $\widetilde{M}_{(5)}=Q_c[\hat{\xi}^\mu_{(5t)}]$.
Note that Eq. (\ref{dM5DKerAdS})
is similar to Eq. (\ref{dM4DKerAdS}) for the 4D Kerr-AdS black hole.
As before,
in order to make the quantity $\widetilde{M}_{(5)}$ coincide with the physically
meaningful mass $M_{(5)}$, a rather effective way is to set $\omega_\phi=-\Omega^\infty_\phi$ and
$\omega_\psi=-\Omega^\infty_\psi$, that is, the angular velocities of the
5D Kerr-AdS black hole vanish at infinity.

A remark is in order here. In the above, we have demonstrated that it is a better choice
to evaluate the mass of the 4D and 5D Kerr-AdS black holes in an asymptotically non-rotating
frame. However, Eq. (\ref{LincomQc12}) allows the possibility to compute their mass in a
rotating frame at infinity when the asymptotic angular velocity
is independent of the mass parameter. For instance, to get the mass $M_{(4)}$ for the 4D
Kerr-AdS black hole (\ref{Metric4DKeAdS}) with $\omega_\phi=0$, one can set that
the variation of the metric only depends on that of the single parameter $m$ and
the Killing vector corresponding to the mass is chosen as the asymptotic
non-rotating timelike Killing vector
$\hat{\xi}^\mu_{(t)}=\xi^\mu_{(t)}+a\ell^2\xi^\mu_{(\phi)} = (-1,0,0,a\ell^2)$.
The former makes the charge $Q_c[\xi^\mu_{(t)}]$ be integrable.
Its integral yields $Q_c[\xi^\mu_{(t)}]=M_{(4)}-a\ell^2J_{(4)}$,
while $Q_c[\xi^\mu_{(\phi)}]=J_{(4)}$. Thus the linear combination of $Q_c[\xi^\mu_{(t)}]$
and $Q_c[\xi^\mu_{(\phi)}]$ gives rise to $Q_c[\hat{\xi}^\mu_{(t)}]=Q_c[\xi^\mu_{(t)}]
+a\ell^2 Q_c[\xi^\mu_{(\phi)}]=M_{(4)}$.
On the other hand, in the case where the parameters
$\omega_\phi,\omega_\psi=0$ within the metric form (\ref{Metricof5DBH}) for the 5D
Kerr-AdS black hole, if the fluctuation of the metric is still determined by the
parameter $m$ rather than the ones $(m,a,b)$, and the timelike Killing vector is set
as $\hat{\xi}^\mu_{(5t)}=\xi^\mu_{(5t)}+a\ell^2\xi^\mu_{(5\phi)}+b\ell^2\xi^\mu_{(5\psi)}$,
one obtains $M_{(5)}$. Besides, in the more general cases, if the Killing vectors
are fixed, the terms proportional to the angular momenta in  Eqs. (\ref{dM4DKerAdS})
and (\ref{dM5DKerAdS}) vanish, yielding the physical mass of the 4D and 5D Kerr-AdS
black holes in the general asymptotically rotating frames.

\section{Mass of Kerr-AdS black holes in arbitrary dimensions}\label{secfour}

In the present section, we deal with the off-shell generalized ADT mass for the
general Kerr-AdS black holes in $D$ $(D\geq 4)$ dimensions by generalizing the
analysis for the 4D and 5D ones. The general Kerr-AdS black holes
in $D=(2N+1)+\epsilon$ dimensions, where $\epsilon=1$ when $D$ is even and
$\epsilon=0$ when $D$ is odd, were constructed in \cite{GLuPPA,GLuPPB}. They possess
$N$ independent rotations in $N$ orthogonal 2-planes, characterized by the parameters
$a_i$'s $(1\leq i\leq N)$ and the azimuthal angles $\phi_i$'s. The metric for the
$D$-dimensional Kerr-AdS black hole is read off as
\bea
ds^2_{(D)}&=&d\bar{s}^2_{(D)}+\frac{2mU}{V(V-2m)}dr^2
+\frac{2m}{U}
\Big(Wdt-\sum_{i=1}^{N}a_i\mu_i^2\frac{d\hat{\phi}_i}{\Xi_i}\Big)^2
\, , \nn \\
d\bar{s}^2_{(D)}&=&-W(1+\ell^2r^2)dt^2
+\sum_{i=1}^{N}\mu_i^2(r^2+a_i^2)\frac{d\hat{\phi}_i^2}{\Xi_i}
+\sum_{i=1}^{N+\epsilon}(r^2+a_i^2)\frac{d\mu_i^2}{\Xi_i}
\nn \\
&&-\frac{\ell^2}{W(1+\ell^2r^2)}
\Big(\sum_{i=1}^{N+\epsilon}\frac{r^2+a_i^2}{\Xi_i}\mu_id\mu_i\Big)^2
+\frac{U}{V}dr^2
\, , \label{DDKerrAdS}
\eea
in which
\bea
U&=&\frac{r^2V}{1+\ell^2r^2}\sum_{i=1}^{N+\epsilon}\frac{\mu_i^2}{r^2+a_i^2}
\, , \quad
V=(r^{\epsilon-2}+\ell^2r^\epsilon)\prod_{i=1}^N(r^2+a_i^2)
\, , \quad
W=\sum_{i=1}^{N+\epsilon}\frac{\mu_i^2}{\Xi_i}
\, , \nn \\
d\hat{\phi}_i&=&d\phi_i-\omega_{(i)}(m,a_j,\ell) dt\, ,
\quad \Xi_i=1-a_i^2\ell^2
\, , \label{WUVdphi}
\eea
and the coordinates $\mu_i$'s obey the constraint $\sum_{i=1}^{N+\epsilon}\mu_i^2=1$.
The metric (\ref{DDKerrAdS}) describes rotating black holes in an asymptotically
rotating frame with angular velocities $\Omega^\infty_{(i)}=\omega_{(i)}$. It becomes the one in
Eq. (4.2) of the paper \cite{GibPPther} when the parameters $\omega_{(i)}$'s vanish,
that is, the Kerr-AdS black holes are non-rotating at infinity.

We turn our attention to evaluating the off-shell generalized ADT mass of the Kerr-AdS
black holes in $D$ dimensions. As is shown in both the 4D and 5D cases, we choose the
asymptotic non-rotating timelike Killing vector associated with the mass as the one
$\hat{\xi}^\mu_{(Dt)}=\xi^\mu_{(Dt)}-\sum_i \Omega^\infty_{(i)}\xi^\mu_{(i)}$,
where the timelike Killing vector $\xi^\mu_{(Dt)}=-\delta^\mu_t$ and the
spacelike Killing vectors $\xi^\mu_{(i)}=\delta^\mu_{\phi_i}$,
while the perturbations of the metric rely on the infinitesimal changes of all the parameters
$(m,a_1, \cdot\cdot\cdot, a_N)$ through
\be
m\rightarrow m+dm \, , \quad
a_i\rightarrow a_i+da_i\quad (i=1, \cdot\cdot\cdot, N) \, . \quad
\ee
Under these conditions, it is proposed that the variation of the off-shell generalized
ADT charge $Q_c[\hat{\xi}^\mu_{(Dt)}]$ for the $D$-dimensional Kerr-AdS black hole described
by Eq. (\ref{DDKerrAdS}) takes the following form
\be
dQ_c[\hat{\xi}^\mu_{(Dt)}]=dM_{(D)}
-\sum_{i=1}^{N}\Big(\frac{\partial\Omega^\infty_{(i)}}{\partial m}dm
+\sum_{j=1}^{N}\frac{\partial\Omega^\infty_{(i)}}{\partial a_j}da_j\Big)J_{(i)}
\, , \label{dMgenKerAdS}
\ee
where
\be
M_{(D)}=\frac{\mathcal{A}_{D-2}}{4\pi}\frac{m}{\prod_j\Xi_j}
\Big(\sum_{i=1}^{N}\frac{1}{\Xi_i}-\frac{1-\epsilon}{2}\Big)
\, , \quad
J_{(i)}=\frac{\mathcal{A}_{D-2}}{4\pi}\frac{ma_i}{\Xi_i\prod_j\Xi_j}
\, . \label{MassAninDdi}
\ee
In the above equation, $\mathcal{A}_{D-2}$ is the volume of the unit $(D-2)$ sphere.
$M_{(D)}$ and $J_{(i)}$ are the physical mass and angular momenta in the literature
\cite{JamsarDker,GibPPther,DerKather,GullTek,BarnichC}. $J_{(i)}$'s, which can result
from the Komar integrals, coincide with the off-shell generalized ADT charges
$Q_c[\xi^\mu_{(i)}]$. It should be emphasized that the following equation
\be
dQ_c[\xi^\mu_{(Dt)}]
=dM_{(D)}+\sum_{i=1}^{N} \Omega^\infty_{(i)} dJ_{(i)}
 \label{dMgenKerAdSt}
\ee
is utilized in order to obtain Eq. (\ref{dMgenKerAdS}). In fact, due to Eq. (\ref{DQc1c2})
in \textbf{Appendix A}, both Eqs. (\ref{dMgenKerAdS}) and (\ref{dMgenKerAdSt}) are
equivalent.

Equation (\ref{dMgenKerAdS}) can be regarded as the generalization of Eqs. (\ref{dM4DKerAdS})
and (\ref{dM5DKerAdS}) in $D$ dimensions. By adopting Eq. (\ref{trcomforADTpogenm}) in
\textbf{Appendix B} to calculate the $(t,r)$ component of the off-shell ADT potential,
this equation has been checked to hold in $D\leq 7$ cases. Furthermore, to make
$Q_c[\hat{\xi}^\mu_{(Dt)}]=M_{(D)}$, one has to cancel the contribution arising from the
variation of the asymptotic velocities $\Omega^\infty_{(i)}$'s. To realize this, a rather
simple method is to perform the coordinate transformations $\phi_i\rightarrow \phi_i+\omega_{(i)} t$
to keep the Kerr-AdS black holes to be static at infinity. This is to say, as long as the
symmetry related to the mass is generated by the usual timelike Killing vector
$\xi^\mu_{(Dt)}$ and the fluctuations of the metric depend on the variation of all
the mass and rotation parameters, only under the condition that these black holes are
asymptotically non-rotating, can the off-shell generalized ADT formalism yield their
physical mass.

For the Kerr-AdS black holes in the asymptotic rotating frame with the
angular velocities $\Omega^\infty_{(i)}$'s irrelevant to the mass parameter $m$,
when the perturbations of the metric are only dependent on the variation of
this parameter, together with the relevant Killing vector given by the asymptotic
non-rotating timelike one $\hat{\xi}^\mu_{(Dt)}$, one is able to obtain the
mass $Q_c[\hat{\xi}^\mu_{(Dt)}]=M_{(D)}$ according to Eq. (\ref{dMgenKerAdS}).

Besides, if the Killing vector $\xi^\mu$ in Eq. (\ref{ADTpotential}) is supposed
to be fixed, that is $\delta\xi^\mu=0$, giving rise to that the conventional ADT potential
$Q_{ADT}^{\mu\nu}$ rather than $\mathcal{Q}_{ADT}^{\mu\nu}$ enters
into the definition of the conserved charge $Q_c$, one observes that the
terms with angular momenta in Eq. (\ref{dMgenKerAdS}) disappear, yielding
$dQ_c[\hat{\xi}^\mu_{(Dt)}]=dM_{(D)}$. Its integral gives the mass of the
$D$-dimensional Kerr-AdS black holes with general asymptotic angular velocities,
which is consistent with the AMD mass \cite{GibPPther,AMDmassA,AMDmassB}.

\section{Summary}\label{secfive}

In this note, we have made use of the off-shell generalized ADT formalism to
compute the mass of the asymptotically-rotating Kerr-AdS black holes
in all dimensions. Without the requirement to fix the Killing vector,
for 4D and 5D Kerr-AdS black holes
in a general asymptotic rotating frame, when the timelike Killing vectors
associated with the mass are chosen as the asymptotic non-rotating timelike
one and the perturbations of the metric are determined by the variation of all
the mass and angular momentum parameters, the conserved charges take the
forms in Eqs. (\ref{dM4DKerAdS}) and (\ref{dM5DKerAdS}) respectively. A similar
equation (\ref{dMgenKerAdS}) holds for the Kerr-AdS black holes in diverse
dimensions. All the equations show that the charges are non-integrable or
unphysical because of the appearance of the general asymptotic angular
velocities. Thus, in order to obtain charges coinciding with the physically
meaningful mass in the literature, the spacetime
has to be non-rotating at infinity. In this sense, all the results support
that the (off-shell) ADT formalism depends on the reference background.

By comparison, in an asymptotic rotating frame with
angular velocities that are only dependent of the rotation parameters, one can obtain
the physical mass of the Kerr-AdS black holes, however, it is required that the
fluctuations of the metric merely rely on the mass parameter and the related Killing
vector is set as the asymptotic non-rotating timelike one. Besides, in
the more general asymptotic rotating frame, one can still get the physical mass if
the asymptotic non-rotating timelike Killing vector is assumed to be fixed.
In the light of the above, we suggest that one had better perform calculations in an asymptotic
non-rotating frame when the off-shell generalized ADT formalism is applied to
compute the mass of asymptotically AdS black holes. This also holds for the
usual ADT formulation and the BBC method since they are essentially equivalent
to the off-shell generalized ADT formalism for the Einstein gravity theory.

It should be emphasized that the cosmological constant is fixed in our calculations so
that it is not involved in the perturbation of the metric. Otherwise, the charges of
the Kerr-AdS black holes are non-integrable. To overcome
this, the off-shell generalized ADT formalism has to be modified by reevaluating the
contribution from the cosmological constant \cite{HJPYLamb}. Accordingly, the mass of
the Kerr-AdS black holes should be reconsidered with the varying cosmological constant.

\section*{Acknowledgments}

This work was supported by the Natural Science Foundation of China under Grant
No. 11505036 and No. 11275157. It was also partially supported by the Technology
Department of Guizhou Province Fund under Grant No. (2016)1104.

\appendix
\section{The linear combination of the off-shell ADT charges}\label{appendA}

Suppose that there exist two Killing vectors $\xi^\mu_{(1)}$ and $\xi^\mu_{(2)}$,
which correspond to the conserved charges $Q_c[\xi^\mu_{(1)}]$ and $Q_c[\xi^\mu_{(2)}]$
respectively. In terms of the formula (\ref{dQdefineAn}), one sees that the variation
of the conserved charge $Q_c[c_1\xi^\mu_{(1)}+c_2\xi^\mu_{(2)}]$ associated with the
linear combination of the two Killing vectors $c_1\xi^\mu_{(1)}+c_2\xi^\mu_{(2)}$
takes the form
\be
\delta Q_c[c_1\xi^\mu_{(1)}+c_2\xi^\mu_{(2)}]=c_1\delta Q_c[\xi^\mu_{(1)}]
+c_2\delta Q_c[\xi^\mu_{(2)}] +(\delta c_1) \mathcal{Q}_{K}[\xi^\mu_{(1)}]
+(\delta c_2) \mathcal{Q}_{K}[\xi^\mu_{(2)}]
\, , \label{DQc1c2}
\ee
where the charge $\mathcal{Q}_{K}$ is defined through the Komar integral
\be
\mathcal{Q}_{K}=\frac{1}{16\pi(D-2)!} \int_{\partial\Sigma} \sqrt{-g}
\nabla^{[\mu}\xi^{\nu]}
\epsilon_{\mu\nu\mu_1\mu_2\cdot\cdot\cdot\mu_{(D-2)}}dx^{\mu_1}\wedge\cdot\cdot\cdot
\wedge dx^{\mu_{(D-2)}}
 \, . \label{KomarInt}
\ee
For another equivalent form of the Komar integral that can be conveniently extended to
the higher-derivative gravity theories see the work \cite{TaTe}.
If the two constants $(c_1,c_2)$ satisfy the restrictions $\delta c_1=0$ and $\delta c_2=0$,
or more generally, the Killing vector $\xi^\mu$ in Eq. (\ref{ADTpotential}) is fixed,
the charge $Q_c[c_1\xi^\mu_{(1)}+c_2\xi^\mu_{(2)}]$ preserves the property of the linearity,
namely,
\be
\delta Q_c[c_1\xi^\mu_{(1)}+c_2\xi^\mu_{(2)}]=c_1 \delta Q_c[\xi^\mu_{(1)}]
+c_2 \delta Q_c[\xi^\mu_{(2)}]
\, . \label{LincomQc12}
\ee

\section{The $(t,r)$ component of the off-shell ADT potential associated with
the Killing vector $\xi^\mu=-\delta^\mu_t$}\label{appendB}

In this appendix, we shall present the explicit form of the $(t,r)$ component
of the off-shell ADT potential for a general metric ansatz that covers
all known stationary and axisymmetric black hole solutions in $D$ dimensions
when the timelike Killing vector is set as $\xi^\mu=-\delta^\mu_t$.

For a black hole in $D$ dimensions, there can exist $N=\frac{D-1-\epsilon}{2}$
independent rotations, where $\epsilon=1$ when $D$ is even and $\epsilon=0$
when $D$ is odd. They correspond to $N$ azimuthal directions $\phi^i$'s.
In the coordinate system $(t,r,\theta^a,\phi^i)$, where $\theta^a$'s denote
the $(D-N-2)$ latitudinal angles, the general metric ansatz describing
all known stationary and axisymmetric black hole solutions can be expressed
as the following form
\be
ds^2=-B\big(dt+Y_id\phi^i\big)^2+Fdr^2+\tilde{g}_{ab}d\theta^ad\theta^b
+\hat{g}_{ij}d\phi^id\phi^j
\, , \label{GenRoMeansa}
\ee
in which all the functions merely depend on the coordinate $r$ and $\theta^a$'s.
In our notation, the indices $a,b,c$ range from 1 to $(D-N-2)$ while the indices
$i,j,k=1,\cdot\cdot\cdot,N$.

By letting the Killing vector $\xi^\mu=-\delta^\mu_t$, in terms of the off-shell
ADT potential in Eq. (\ref{ADTpotential}), its $(t,r)$ component for the metric
(\ref{GenRoMeansa}), defined by $\mathbf{Q}=Q_{ADT}^{tr}[-\delta^\mu_t]$,
is given by
\bea
\mathbf{Q}&=&\frac{1}{4BF^2}\big[
2F\delta(\alpha B)-\alpha\delta (BF)
+2FB^2\hat{g}^{ij}\dot{Y}_i\delta Y_j
+\tilde{P}+\hat{P}\big]
\, , \nn \\
 \tilde{P}&=&
F\big(\dot{B}+\alpha B\big)\tilde{g}^{ab}\tilde{h}_{ab}
-2BF\tilde{g}^{ab}\dot{\tilde{h}}_{ab}
-B\dot{\tilde{g}}^{ab}\delta\big(F\tilde{g}_{ab}\big)
\, , \nn \\
\hat{P}&=&
F\big(\dot{B}+\alpha B\big)\hat{g}^{ij}\hat{h}_{ij}
-2BF\hat{g}^{ij}\dot{\hat{h}}_{ij}
-B\dot{\hat{g}}^{ij}\delta\big(F\hat{g}_{ij}\big)
\, . \label{trcomforADTpogenm}
\eea
In the above equation,
\be
\tilde{h}_{ab}=\delta\tilde{g}_{ab}\, , \quad
\hat{h}_{ij}=\delta\hat{g}_{ij}\, , \quad
\alpha=B\hat{g}^{ij}Y_i\dot{Y}_j  \, ,
\ee
and the dot `` $\dot{}$ " denotes the partial derivative with respect to the coordinate
$r$, for instance, $\dot{Y}_i=\partial_r Y_i$, $\dot{\hat{h}}_{ij}=\partial_r \hat{h}_{ij}$,
$\dot{\hat{g}}^{ij}=\partial_r \hat{g}^{ij}$
and so on. Equation (\ref{trcomforADTpogenm}) simplifies the calculations of the ADT
potential quite drastically and it can be widely applied to the Einstein gravity theory
coupled with matter fields or not.


\end{document}